\documentclass[aip,jcp,preprint]{revtex4-1}
\def\bv{{\bf v}}
\def\br{{\bf r}}
\newcommand{\be}{\begin{equation}}
\newcommand{\ee}{\end{equation}}
\newcommand{\beq}{\begin{eqnarray}}
\newcommand{\eeq}{\end{eqnarray}}
\newcommand{\VEC}[1]{{\bf #1}}
\newcommand{\T}{\overline{T}}
\newcommand{\ff}{f_{2}}
\newcommand{\fff}{f_{3}}
\newcommand{\derpar}[2]{\frac{\partial #1}{\partial #2}}

\newcommand{\s}{\hat{\mbox{\boldmath $\sigma$\unboldmath}}}
\newcommand{\si}{{\mbox{\boldmath $\sigma$\unboldmath}}}
\newcommand{\vs}[2]{\VEC{v}_{#1#2}\cdot\s}
\usepackage{amsmath}
\usepackage{graphicx}

\begin{document}

\title{Prediction of a Structural Transition in the Hard Disk Fluid}
\author{Jaros\l{}aw Piasecki}
\email{Jaroslaw.Piasecki@fuw.edu.pl} 
\author{Piotr Szymczak}
\affiliation{Institute of Theoretical Physics, University of Warsaw,
 Hoza 69, 00-681 Warsaw, Poland} 
\author{John J. Kozak}
\affiliation{DePaul University, 243 South Wabash Avenue, Chicago, Illinois 60604-2301, U.S.A.}

\begin{abstract}
Starting from the second equilibrium equation in the BBGKY hierarchy under the Kirkwood superposition closure, we implement a new method for studying the asymptotic decay of correlations in the hard disk fluid in the high density regime.  From our analysis and complementary numerical studies, we  find  that exponentially damped oscillations can occur only up to a packing fraction $\eta^* \sim 0.718$,  a value which is in substantial agreement with the packing fraction,  $\eta \sim 0.723$,  believed to characterize the transition from the ordered solid phase to a dense fluid phase, as inferred from Mak's Monte Carlo simulations [Phys. Rev. E {\bf 73}, 065104 (2006)]. We next show that the same method of analysis predicts that exponential damping of oscillations in the hard sphere fluid becomes impossible when 
$\lambda = 4n\pi \sigma^3 [1 + H(1)] \geq 34.81$, where $H(1)$ is the contact value of the correlation function, n is the number density and $\sigma$ is the sphere diameter, in exact agreement with the condition, 
$\lambda \geq 34.8$,  first reported in a numerical study of the Kirkwood equation by  Kirkwood et al. [J. Chem. Phys. {\bf 18}, 1040 (1950)].  Finally, we show that our method confirms the absence of any structural transition in hard rods for the entire range of densities below close packing. 
\end{abstract}

\maketitle

\section{Introduction}

The second equilibrium equation in the BBGKY hierarchy establishes an exact relation between the pair and triplet number density.  Invoking the Kirkwood superposition approximation yields a nonlinear integral equation for the pair correlation function \cite{r1,r2}.  Interest in studying the analytic and numerical properties of the resulting Yvon-Born-Green and/or Kirkwood equation began with Kirkwood and coworkers \cite{r3,r4,r5,r6},  and continues to the present day \cite{r7}.  Of particular interest is whether the closed equation provides an essentially correct description of the fluid phase, and whether a (possible) change in the analytic character of the solutions signals a change from the fluid phase to a solid phase.  It is to the latter question that the methods of the present contribution are directed.

One approach to explore analytically the possibility of a phase transition from the fluid phase to the solid phase is to mobilize the theory of nonlinear integral equations, focusing on theorems which establish the necessary and sufficient conditions for the existence and uniqueness of solutions, and bifurcation points \cite{r8,r9,r10,r11,r12}.  An alternative approach is to introduce a moment expansion by means of which the YBG equation can be cast into a nonlinear differential equation which may be used to analyze long-range correlations \cite{r13,r14}.  The present contribution is centered on a new method of studying the asymptotic decay of correlations, first introduced in Ref. 15 for the hard sphere fluid.  

In this paper, we focus on the hard disk fluid.  The method leads to the prediction of a structural transition in both the hard sphere and hard disk fluids, and no transition in the hard rod system (as must be the case).  We shall show that the values of packing fractions at which the predicted transitions occur are in agreement with estimates derived from numerical solution of the Kirkwood equation \cite{r6} and recent Monte Carlo simulations \cite{r16}.

\section{The second equilibrium hierarchy equation for hard disks}
We consider a gas of hard disks of diameter $\sigma$ at thermal equilirium with constant number density $n$ and temperature $T$.

Let $n_{2}(r_{12})$ denote the number density of pairs of particles situated at distance $r_{12}=|\VEC{r}_{1}-\VEC{r}_{2}|$. 
Using the fact that  $n_{2}(r_{12})=0$ for $r_{12}<\sigma$ we introduce a dimensionless function $y_{2}(r_{12})$ defined by 
\be
\label{n2}
n_{2}(r_{12}) = n^{2}\,\theta(r_{12}-\sigma)y_{2}(r_{12}) 
\ee
 where $\theta$ is a unit step function.

We assume that $n_{2}(r_{12}) \to n^{2}$ when $r_{12}\to\infty$. 
The two-particle dimensionless correlation function $h_{2}(r_{12})$ is then defined by the cluster decomposition $y_{2}(r_{12})=1 + h_{2}(r_{12})$, so that
\be
\label{h2}
\frac{n_{2}(r_{12})}{n^{2}} = \theta(r_{12}-\sigma)[ 1 + h_{2}(r_{12}) ] 
\ee
$h_{2}(r_{12})$  is thus supposed to satisfy the asymptotic condition
\be
\label{as}
 \lim_{r_{12}\to\infty} h_{2}(r_{12}) = 0  
\ee

The second equilibrium Yvon-Born-Green (YBG) hierarchy equation establishes an 
exact relation between 
$n_{2}(r_{12})$ and the reduced three-particle number density $n_{3}(\br_{1},\br_{2},\br_{3})$.
Introducing the excluded volume factor we write the three-particle density as
\be
\label{n3}
n_{3}(\br_{1},\br_{2},\br_{3})= \theta(r_{12}-\sigma) \theta(r_{13}-\sigma) \theta(r_{23}-\sigma)n^{3}y_{3}(r_{12},r_{13},r_{23})
\ee
The function $n_{3}$ depends only on the distances $r_{ij} =|\br_{ij}|= |\br_{i}-\br_{j}|$, and is a symmetric function of the three variables $r_{12}, r_{13}, r_{23}$.
In the case of hard disks $y_{2}$ is related to $y_{3}$ through the second YBG hierarchy equation (see Appendix A)
\be
\label{YBGy}
\frac{d}{dr}y_{2}(r) = n\sigma \int d\s ( \hat{\VEC{r}}\cdot \s )  \theta( |\VEC{r}-\sigma\s | - \sigma) 
 y_{3}(r, \sigma, |\br-\sigma\s|)
\ee  
where  $\s $ and $\hat{\VEC{r}}$ are unit vectors $|\hat{\boldsymbol{\sigma}}|=|\hat{\boldsymbol{r}}| =1$. Putting $\hat{\VEC{r}}\cdot \s = \cos\phi$ we rewrite (\ref{YBGy}) in an explicit form
\be
\label{YBGyy}
\frac{d}{dr}y_{2}(r) = n\sigma \int_{0}^{2\pi} d\phi \cos\phi \,
 y_{3}(r, \sigma, \sqrt{r^{2}-2r\sigma\cos\phi+\sigma^{2}}) \theta( r - 2\sigma\cos\phi)
\ee
When writing (\ref{YBGyy}) the equality
\[ \theta( |\VEC{r}-\sigma\s |-\sigma ) = \theta( r - 2\sigma\cos\phi)  \]
has been used.

The rigorous relation (\ref{YBGyy}), valid for $r>\sigma$,  will be the starting point for subsequent considerations.

\section{The Kirkwood superposition approximation}

The Kirkwood superposition approximation consists in replacing in equation (\ref{YBGyy}) the three-particle density
 by the product of two-particle densities corresponding to three different pairs of particles
\be
\label{superposition}
y_{3}(r, \sigma, |\br-\sigma\s|)\;\;\;\; \to \;\;\;\;  y_{2}(r) y_{2}(\sigma) y_{2}( |\br-\sigma\s|)
\ee
Adopting (\ref{superposition}) leads to a closed equation
\be
\label{Ky}
\frac{d}{dr}y_{2}(r) = n\sigma y_{2}(r) y_{2}(\sigma)\int_{0}^{2\pi} d\phi \cos\phi \,y_{2}(\sqrt{r^{2}-2r\sigma\cos\phi+\sigma^{2}})
  \theta( r - 2\sigma\cos\phi)
\ee
It is convenient to rewrite (\ref{Ky}) using the dimensionless distance $x = r/\sigma$. Denoting by $Y(x)$ the function 
\be
\label{Y}
Y(x) = y_{2}(x\sigma)
\ee
we find that it satisfies the non-linear equation
\be
\label{KY}
\frac{d}{dx} \ln Y(x) = n\sigma^{2}  Y(1)\,\int_{0}^{2\pi} d\phi \cos\phi \, Y(\sqrt{x^{2}-2x\cos\phi+1})
  \theta( x - 2\cos\phi)
\ee
valid in the region $x \ge 1$.
Equation (\ref{KY}) represents the closure of the YBG hierarchy corresponding to the superposition approximation.

 Our aim is to derive from (\ref{KY})  the equivalent integral equation satisfied by the dimensionless correlation function
\be
\label{H}
H(x) = Y(x) - 1
\ee
To this end we insert (\ref{H}) into (\ref{KY}) finding
\be
\label{KH1}
 \frac{d}{dx}\ln [1+H(x)]=n\sigma^{2}[  1+H(1) ]\left\{ \,\int_{0}^{2\pi} d\phi \cos\phi \, \theta( x - 2\cos\phi) \right.
\ee
\[ \left. + \int_{0}^{2\pi} d\phi \cos\phi \, H(\sqrt{x^{2}-2x\cos\phi+1})\theta( x - 2\cos\phi)  \right\}\]

In order to derive an integral equation for $H(x)$ we integrate (\ref{KH1}) over the spatial interval $(x,\infty )$. Using the formulae
\be
\label{angular}
\int_{0}^{2\pi }d\phi \, \theta(x-2\cos\phi ) \cos\phi = -2\theta(2-x) \sqrt{1-\left(\frac{x}{2}\right)^2}   
\ee
\be
\label{f2}
\int_{x}^{\infty}dz\, \theta(2-z) \sqrt{1-\left(\frac{z}{2}\right)^2} =\theta(2-x) \left[ -\frac{x}{2}\sqrt{1-\left(\frac{x}{2}\right)^2}+
\arccos\frac{x}{2}\right]
\ee
we get
\[ \ln [1+H(x)] =2n\sigma^{2} [  1+H(1)] [I^*(x) + I^{**}(x)]  \]
where
\be
\label{i1}
I^*(x) = \theta(2-x) \left[ - \frac{x}{2}\sqrt{1-\left(\frac{x}{2}\right)^2} + \arccos\frac{x}{2}\right]
\ee
and
\be
\label{i2}
 I^{**}(x)= - \frac{1}{2} \int_{x}^{\infty} dz\,\int_{0}^{2\pi }d\phi \, \cos\phi\, \theta( z-2\cos\phi )H(\sqrt{z^{2}+1-2z\cos\phi})   
\ee
It turns out that the angular integration in the formula for $I^{**}(x)$ can be explicitly performed (the calculation is presented in Appendix B). One eventually finds
\be
\label{**}
 I^{**}(x)= - \int_{x-1}^{x+1}ds\,\theta(s-1)\,  s H(s) \arccos\left( \frac{x^{2}+s^{2}-1}{2xs} \right)  
\ee
In this way we arrive at an integral equation
\be
\label{KHH}
 \ln [1+H(x)] = 2n\sigma^{2}[1+H(1)] \left\{ \theta(2-x) \left[ - \frac{x}{2}\sqrt{1-\left(\frac{x}{2}\right)^2} + 
\arccos\frac{x}{2}\right] \right.
\ee
\[  \left. -  \int_{x-1}^{x+1}ds\,\theta(s-1)\,  s H(s) \arccos\left( \frac{x^{2}+s^{2}-1}{2xs} \right) \right\}\]
representing the superposition closure for the correlation function $H(x)$.

Our method of determining $H(x)$  is based on the fact that the solution 
of Eq. (\ref{KHH})  satisfying the boundary condition  $ \lim_{x\to\infty}H(x)=0$  can be obtained numerically by iterations.  Note that we can rewrite Eq.~(\ref{KHH}) in the following form
\be
\label{KHH2}
H(x) = {\cal{L}}(H(x))
\ee
where $\cal{L}$ is the integral operator given by
\begin{multline}
\label{KHH3}
{\cal{L}}(H(x)) \equiv -1 + \exp\left( 2n\sigma^{2}[1+H(1)] \left\{ \theta(2-x) \left[ - \frac{x}{2}\sqrt{1-\left(\frac{x}{2}\right)^2} + 
\arccos\frac{x}{2}\right] \right. \right. \\ \left. \left.
 -  \int_{x-1}^{x+1}ds\,\theta(s-1)\,  s H(s) \arccos\left( \frac{x^{2}+s^{2}-1}{2xs} \right) \right\} \right) 
\end{multline}
 The above integral equation for $H(x)$ was solved by a standard Neumann method with  succesive over-relaxation~\cite{r18}. The iterative solutions are then given by
\begin{equation}
H_n = (1-\alpha)H_{n-1} + \alpha {\cal{L}}(H_{n-1})
\end{equation}

The relaxation parameter $\alpha$ was taken to be 0.25. The iterations were continued until successive values of $H(x=0)$ differed by less than $\epsilon=10^{-5}$, except in the vicinity of the threshold surface fraction $\xi^*$ (see Sec.~\ref{pred}), where the convergence was slow and the iterations were discontinued at $\epsilon=10^{-2}$.

\begin{figure}
\center\includegraphics[width=5in]{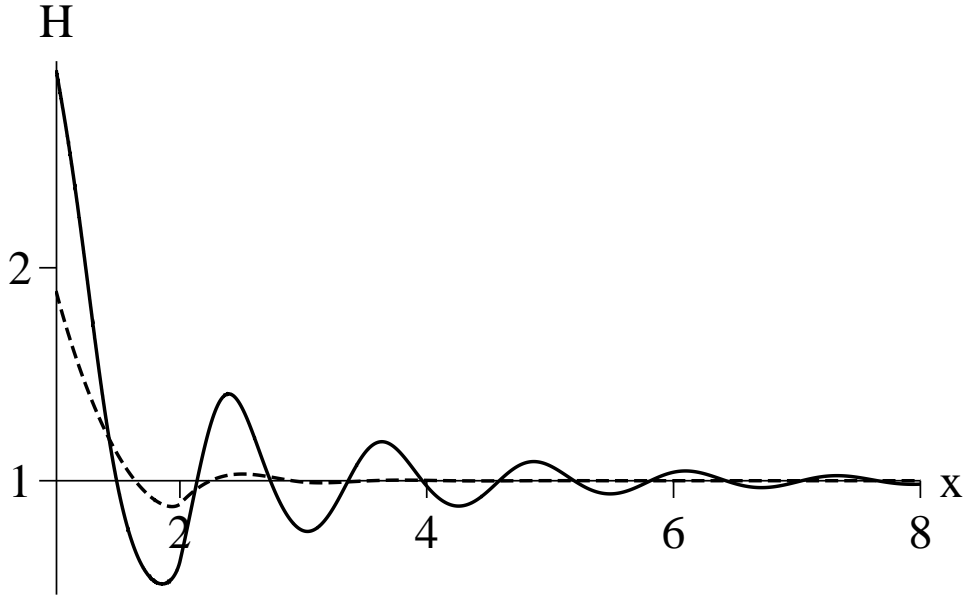}
\caption{Pair correlation function $H(x)$ for the surface fraction $\xi=n \pi \sigma^2 /4 = 0.35$ (dashed line) and $\xi=0.59$ (solid line).}\label{correl}
\end{figure}

Examples of the correlation functions obtained in this way are presented in Fig.~\ref{correl}. As it is seen, the decay of $H(x)$ becomes slower as the surface fraction is increased, and a pronounced peak structure appears.

\section{Linearization of Kirkwood's equation: the ring approximation}

Before continuing the analysis based on equation (\ref{KHH}) let us make a comment on the relationship between the superposition approximation and the  ring approximation, well known from the kinetic theory (see \cite{r15} and references given therein).

 Originally, the ring approximation was applied to the study of long wavelength hydrodynamic phenomena, and was defined by neglecting in the second equation of the dynamical BBGKY hierarchy all contributions from the three-particle correlations.
 The three-particle correlation function $h_{3}$ is defined by the cluster decomposition of the density $y_{3}$ 
\be
\label{corr3}
 y_{3}(r_{12},r_{13},r_{23}) = 1 + h_{2}(r_{12}) + h_{2}(r_{13}) + h_{2}(r_{23}) + h_{3}(r_{12},r_{13},r_{23}) 
\ee
Neglecting $h_{3}$ is thus equivalent to the approximation
\be
\label{ringapp}
 y_{3}(r_{12},r_{13},r_{23}) \simeq 1 + h_{2}(r_{12}) + h_{2}(r_{13}) + h_{2}(r_{23}) 
\ee
 It is important here to note that while rejecting in the cluster decomposition (\ref{corr3}) three-particle correlations we nevertheless retain in the hierarchy equation the full excluded volume factor represented by the product of unit step-functions (see definition (\ref{n3})). In fact, this factor represents the exact lowest order term in the density expansion of the
 three-particle number density, and is of fundamental importance for the correct description of hard disks at low densities. 
 
Comparing (\ref{ringapp}) with the superposition formula
\be
\label{supapp}
y_{3}(r_{12},r_{13},r_{23}) \simeq [1 + h_{2}(r_{12})][1 + h_{2}(r_{13})][1 + h_{2}(r_{23})] 
\ee
we see that  the ring approximation corresponds exactly to the linearization of the Kirkwood theory in $h_2$. 

The linearized form of equation (\ref{KH1}) reads
\be
\label{re5}
 \frac{d}{dx}H(x)=n\sigma^{2}\left\{ -2\theta(2-x) \sqrt{1-\left(\frac{x}{2}\right)^2} [  1+H(1) +H(x)] \right.
\ee  
\[ \left.  + \int_{0}^{2\pi }d\phi \, \cos\phi\, \theta( x-2\cos\phi )H(\sqrt{x^{2}+1-2x\cos\phi}) \right\} \]

In order to derive an integral equation for $H(x)$ we proceed as before by integrating (\ref{re5}) over the spatial interval $(x,\infty )$.
The result reads
\[ H(x) =2n\sigma^{2} [J^*(x) + J^{**}(x)]  \]
where
\be
J^*(x) = \theta(2-x)\left\{ \left[ - \frac{x}{2}\sqrt{1-\left(\frac{x}{2}\right)^2} + \arccos\frac{x}{2}\right][  1+H(1)] +
\int_{x}^{2} dz\, \sqrt{1-\left(\frac{z}{2}\right)^2}H(z)] \right\}
\ee
and
\[ J^{**}(x)= - \frac{1}{2} \int_{x}^{\infty} dz\,\int_{0}^{2\pi }d\phi \, \cos\phi\, \theta( z-2\cos\phi )H(\sqrt{z^{2}+1-2z\cos\phi})  =
 I^{**}(x) \]
(see equation (\ref{i2})).
Using again the calculation presented in Appendix B we eventually find
\be
\label{***}
 J^{**}(x)= - \int_{x-1}^{x+1}ds\,\theta(s-1)\,  s H(s) \arccos\left( \frac{x^{2}+s^{2}-1}{2xs} \right)  
\ee
In this way we arrive at an integral equation
\be
\label{Ring}
H(x) =2n\sigma^{2} \theta(2-x)\left\{ \left[ - \frac{x}{2}\sqrt{1-\left(\frac{x}{2}\right)^2} + \arccos\frac{x}{2}\right][  1+H(1)] +
\int_{x}^{2} dz\, \sqrt{1-\left(\frac{z}{2}\right)^2}H(z)] \right\}
\ee
\[ - 2n\sigma^{2}  \int_{x-1}^{x+1}ds\,\theta(s-1)\,  s H(s) \arccos\left( \frac{x^{2}+s^{2}-1}{2xs} \right)   \]
representing the ring approximation.  
Equation (\ref{Ring}) can again be solved numerically by iterations, and is expected to yield relevant results in the low density regime. 

\begin{figure}
\center\includegraphics[width=5in]{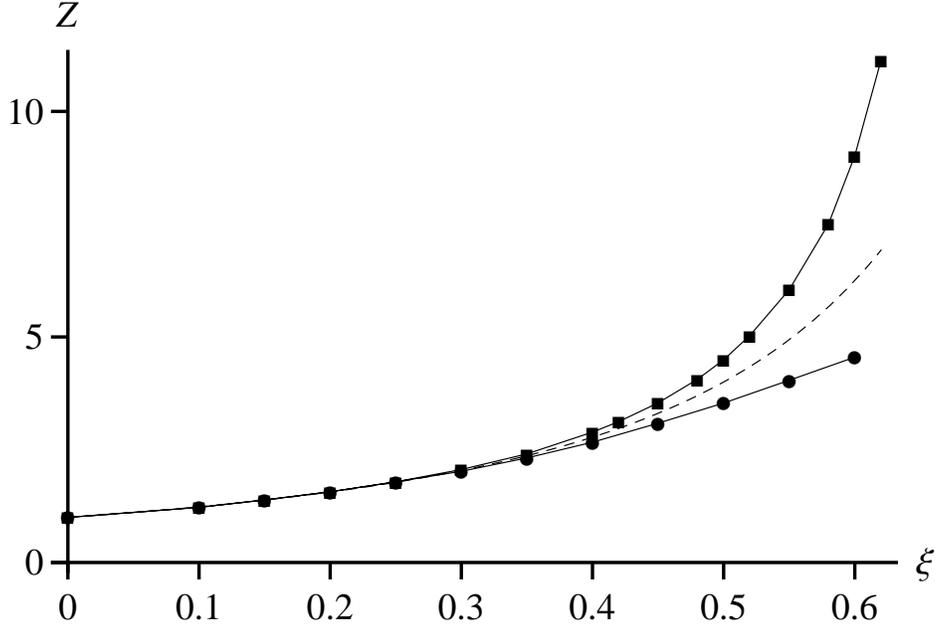}
\caption{Compressibility factor  $Z$ versus the surface fraction $\xi$ for the full Kirkwood approximation (circles), the ring approximation (squares), and the scaled particle theory (Eq.~(\ref{spt}), dashed line)}\label{pressure}
\end{figure}

The comparison of the results obtained with the full Kirkwood approximation and its linearization is presented in  Fig.~\ref{pressure}.
Defining the surface fraction $\xi$ as
\be
\label{surface}
\xi = n\pi \frac{\sigma^2}{4}
\ee 
we plot here the compressibility factor $Z(\xi)$, given by
\begin{equation}
Z(\xi)=\frac{p}{n kT} = 1 + \frac{\pi }{2} n\sigma^{2} (1+H(1)) =  1 + 2 \xi (1+H(1)),
\end{equation}
where the contact value $H(1)$ is also a function of $\xi$.
For comparison, we include here $Z(\xi)$ dependence as predicted by the scaled particle theory (SPT)~\cite{r19}
\begin{equation}
Z_{SPT}(\xi)=\frac{1}{(1-\xi)^2} 
\label{spt}
\end{equation}
As it is seen, the iteration results agree with the SPT predictions up to approximately $\xi=0.4$. For larger packing fractions, the Kirkwood approximation tends to underestimate the compressibility factor, whereas ring approximation  overestimates it.

\section{Asymptotic decay of correlations: predicting a structural transition}

\subsection{Breakdown of the method used for attractive interactions}

The fundamental information concerning the internal structure of the system is contained in the spatial dependence of correlations. In  particular, the law governing the asymptotic vanishing of correlations is of primary importance.

In order to determine the behavior of H(x) at large distances, it seems natural to
follow the moment analysis presented in Refs. 13 and 14.  The calculation would proceed as
follows.

For $x>2$, the Kirkwod equation (\ref{KH1}) can be conveniently written as
\be
\label{x>2}
\frac{d}{dx}\ln [1+ H(x) ] = n\sigma^{2}\,[1+ H(1)]\, \int d\s ( \hat{\VEC{x}}\cdot \s ) H( |\VEC{x}-\s | )
\ee
When $x\gg 1$, the power series expansion of $H( |\VEC{x}-\s | )$ around the point $|\VEC{x}| = x$ yields nonzero contributions only from terms involving odd powers of $\cos\phi = ( \hat{\VEC{x}}\cdot \s )$. The calculation up to the third derivative of $H$ yields the expansion
\be
[\ln [1+ H(x) ]' = n\sigma^{2}\,[1+ H(1)]\, \int_{0}^{2\pi} \cos\phi   \{ - \cos\phi \, H'(x) 
\ee
\[  -\frac{1}{2}[ \cos\phi  - (\cos\phi )^{3}]\left( \frac{H'(x)}{x}\right)' - \frac{1}{6}( \cos\phi )^{3}H'''(x)  +  ...  \} \]
where  $'$ denotes the derivative with respect to $x$. 

Using the boundary condition $\lim_{x\to\infty}H(x)=0$ together with the equalities
\[  \int_0^{2 \pi} d\phi ( \cos\phi  )^{2} = \pi, \;\;\;\;      \int_0^{2 \pi} d\phi ( \cos\phi  )^{4}= \frac{3}{4}\pi , \] 
and adopting for large distances the asymptotic formula
\[  \ln [1+ H(x) ] \sim H(x) \]
we find a linear differential equation of the form
\be
\label{bessel}
H''(x) + \frac{1}{x}H'(x) + \alpha^{2} H = 0
\ee
where 
 \[ \alpha^{2} = 8( 1 + \frac{1}{n\pi \sigma^{2}[1+ H(1)]})  > 0 \]
But (\ref{bessel}) is the equation for the Bessel function $J_{0}(\alpha x)$. The solution of (\ref{bessel}) is thus an oscillating function whose amplitude decays as $1/\sqrt{x}$ . 

Unfortunately, the above result is in disagreement with numerical predictions of 
exponentially damped oscillations. Use of the moment expansion as developed in Ref.~13  
leads to erroneous results.
Clearly one has to take into account the whole infinite series in the expansion of $H( |\VEC{x}-\s |)$ to get reliable predictions.
We have thus to give up this kind of expansion, and look for a different approach.

In broad outline, the failure of  the moment expansion developed in Refs.~13 and 14 to  
describe the structural transition in the hard disk fluid can be traced to the nature of the 
governing intermolecular potential.  The studies \cite{r13,r14} deal with the description of 
correlations in the vicinity of the liquid-vapor critical point, where attractive forces
play a dominant  role.  Using the moment expansion, one recovers at large distances
the classical Ornstein-Zernike formula.  The present study deals with the fusion 
transition, where short-range, repulsive forces play the critical role.   The new method 
introduced here (see Sec.~\ref{pred}) effectively accounts for the difference in the potential
governing these two transitions and, anticipating our later results, leads to an analytic
prediction of the packing fraction at  which a structural transition occurs in the hard
disk (and hard sphere) fluid.

\subsection{Prediction of a structural transition}\label{pred}

Let us consider again the region $x>2$ where the equation (\ref{x>2}) holds. As $\lim_{x\to\infty}H(x)=0$, we can replace in (\ref{x>2}) the function $\ln [1+ H(x) ]$ by $H(x)$, and consider the equation
\be
\label{AS1}
\frac{d}{dx}H(x)  =  \frac{A}{2\pi} \, \int_{0}^{2\pi} d\phi \cos\phi\, H( \sqrt{x^2 - 2x\cos\phi +1} )
\ee
where 
\[ A = 2\pi n\sigma^{2}\,[1+ H(1)]  \]
We then use the expansion
\be
\label{AS2}
\sqrt{x^2 - 2x\cos\phi +1} = x - \cos\phi + \frac{\sin\phi^2}{2x} + ...
\ee
to arrive at the equation 
\be
\label{AS3}
\frac{d}{dx}H(x)  = \frac{A}{2\pi} \int_{0}^{2\pi} d\phi \cos\phi\, H( x-\cos\phi ),
\ee
valid for $x \gg 1$.

\begin{figure}
\center\includegraphics[width=5in]{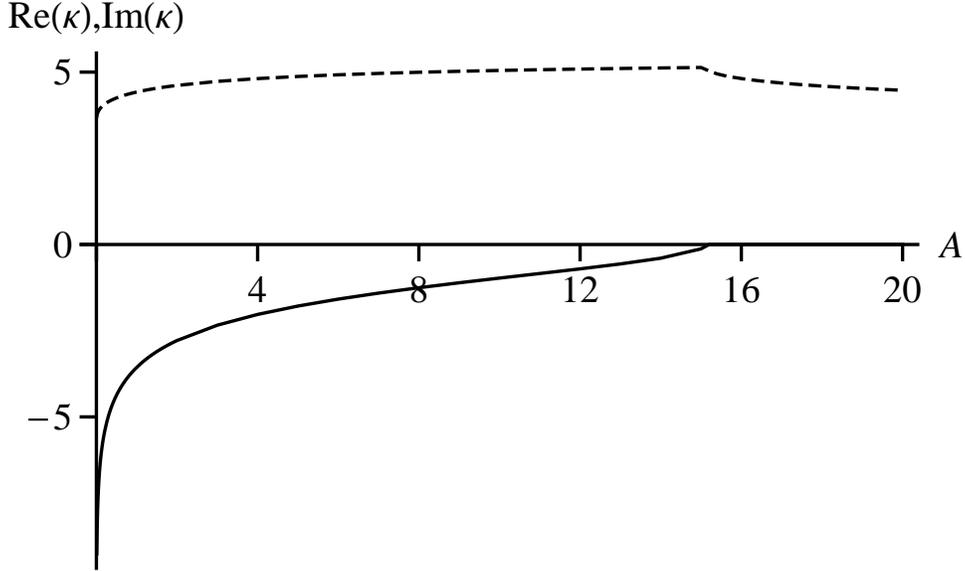}
\caption{The real part (solid line) and imaginary part (dashed line) of the root of Eq.~(\ref{zequ}) corresponding to the slowest decaying mode.}\label{2d}
\end{figure}

\begin{figure}
\center\includegraphics[width=5in]{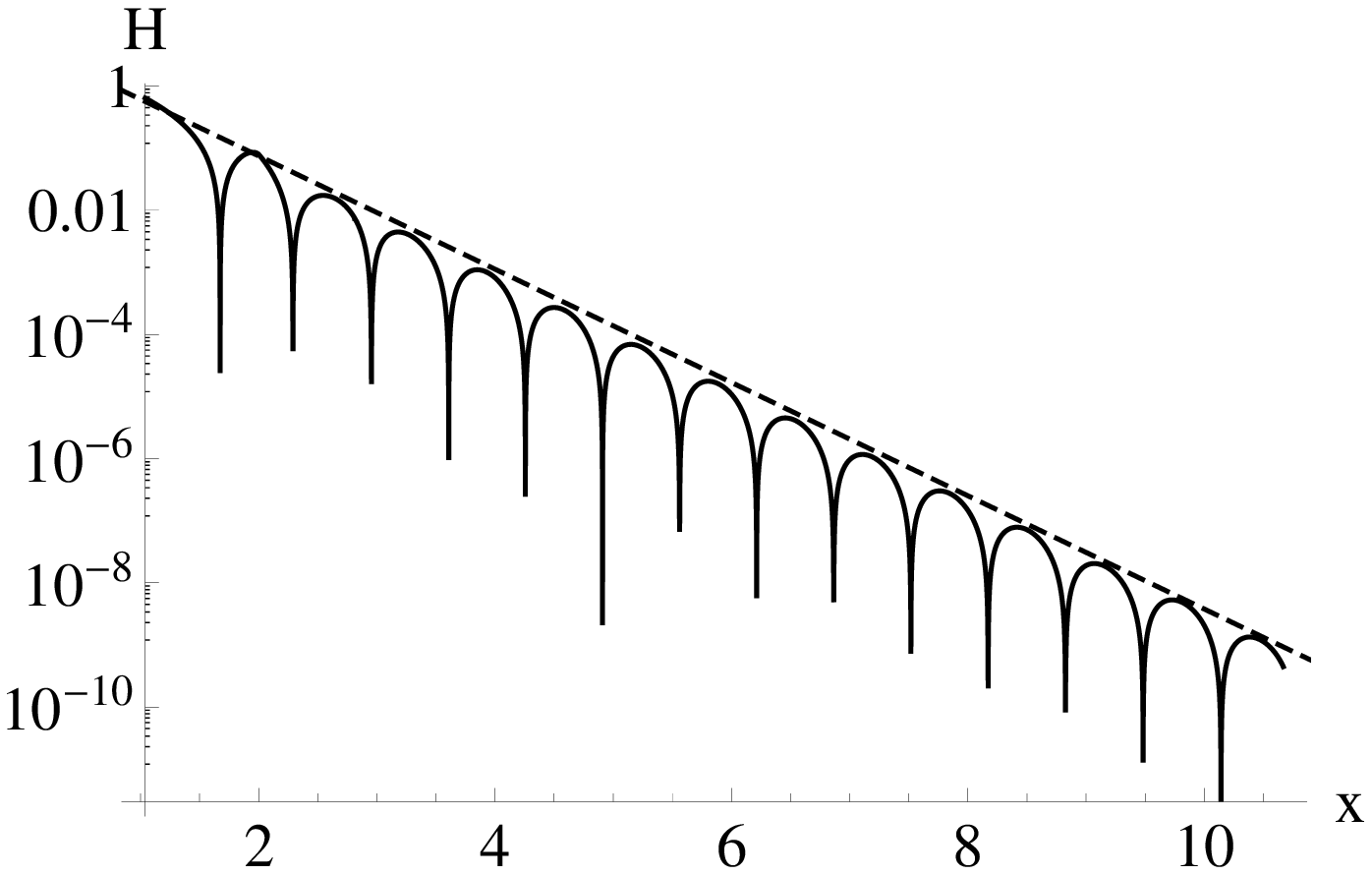}
\caption{The absolute value of $H(x)$ for $\xi=0.3$ together with a corresponding fit of the form $a e^{\kappa' x}$ with $\kappa'=-2.1$.}\label{decay}
\end{figure}

In order to determine the large $x$ behavior of correlations we have thus to analyze the solution of (\ref{AS3}).
We notice that {\it all derivatives of} $H(x)$ {\it satisfy the same equation}. It is thus natural to consider $H(x)$ as a linear combination of exponential modes $\exp(\kappa x)$, where $\kappa$ is a complex number. The function $\exp(\kappa x)$ satisfies (\ref{AS3}) provided 
$\kappa$ solves the equation
\be
\label{zequ}
\kappa = \frac{A}{2\pi} \int_{0}^{2\pi} d\phi \cos\phi\, \exp(-\kappa \cos\phi) = -  A I_{1}(\kappa)
\ee
where $I_{1}(\kappa)$ is a modified Bessel function.
The physically acceptable solutions $\kappa = a + ib$ are those with negative real part $a<0$ which assures exponential damping of oscillations. The first root (with the smallest absolute value of $a$), corresponding to the slowest decay of the correlation function is shown in Fig.~\ref{2d}. The values of $\kappa (\xi)$ predicted with the use of Eq.~\eqref{zequ} are in good agreement with the decay of the amplitude of $H(x)$ determined by the iterative solution of integral equation~(\ref{KHH2}). Plotting the absolute value of $H(x)$ on a logarithmic plot, and fitting it by the single mode $\alpha e^{\kappa' x}$, we obtain values of $\kappa'$ which are slightly below those obtained by solving Eq.~\eqref{zequ}. For example, for $\xi=0.3$ we get $\kappa'\approx-2.1$ ({\it cf.} Fig.~\ref{decay}), whereas the corresponding value of $\kappa$ for that surface fraction is $\kappa \approx -2.02$. Similarly, for $\xi=0.55$ we get respectively $\kappa'\approx-0.8$ and $\kappa \approx-0.7$.  The fact that the values of $\kappa'$ remain slightly below those of $\kappa$ can be understood by noting that $\kappa$ corresponds to the slowest decaying mode, whereas in the numerical data on $H(x)$ we also see nonzero contributions from other, faster decaying modes.

\begin{figure}
\center\includegraphics[width=5in]{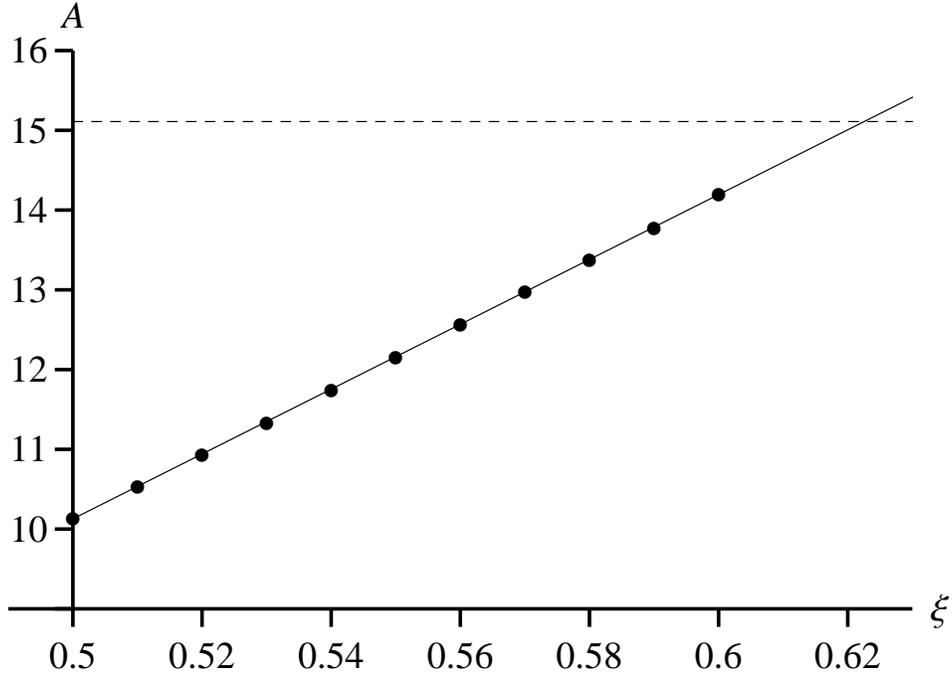}
\caption{The extrapolation of $A(\xi)$ dependence. The points correspond to the values of $A$ obtained from the iterative procedure, the dashed line is given by $A=A^*\approx 15.12$ }\label{extrap}
\end{figure}

An interesting feature of the $\kappa(\xi)$ dependence presented in Fig.~\ref{2d} is the fact that the root becomes purely imaginary at $A= A^* \approx 15.1$. 
This point can be made more precise by analyzing when the equation (\ref{zequ}) acquires a purely imaginary solution $\kappa=ib$. As 
$I_{1}(ib) = i J_{1}(b)$, (\ref{zequ}) implies the condition
\be
\label{condition1}
\frac{J_{1}(b)}{b} = \frac{1}{2}[ J_{0}(b) + J_{2}(b)  ]= -\frac{1}{ A}
\ee
where the first equality follows from the recurrence relation between Bessel functions $J_{n}$.
The absolute minimum of the sum of Bessel functions $[ J_{0}(b) + J_{2}(b)]$ equals $-0.1323 $. The necessary condition for the disappearance of damping has thus the form
\be
\label{condition2}
 \frac{2}{A} \leq 0.1323  ,  \;\;\;\;\; {\rm or}  \;\;\;\;\;  A = 2\pi  n\sigma^{2}\,[1+ H(1)] \geq A^* = 15.1171... \simeq 15.12
\ee
The above inequality shows that the nature of correlations could change for sufficiently high values of the surface fraction $\xi$ [Eq.(\ref{surface})] occupied by hard disks. The determination of the value of $\xi^* $ corresponding to the equality 
\[ A^* = 8 \xi^* \,[1+ H(x=1;\xi^*)] = 15.12 \] 
requires the knowledge of the contact value $H(1)$ as function of the surface fraction. We have studied this question numerically. 
Accurate estimation of the precise value of $\xi^*$ corresponding to $A^*$ is difficult because as we approach  $\xi^*$ the iteration procedure demands an increasingly larger number of iterations to converge to a solution with the required accuracy.
Additionally, a computational domain over which the solution is sought must also be progressively extended as we approach $\xi^*$, since the decay of $H(x)$ is very weak there. 
We estimated the value of $\xi^*$ by calculating $A(\xi)$ for several values of $\xi$ in the range $0.5 \leq \xi \leq 0.6$ and then extrapolating to larger values of $\xi$, as illustrated in Fig.~\ref{extrap}. In this way, we obtain the estimate of $\xi^* \approx 0.622$. 

Let us close this section with a comment on the ring approximation. In order to find the asymptotic behavior of correlations satisfying equation (\ref{Ring}) it is sufficient to consider the linearized version of equation (\ref{AS1}). But this is equivalent to the replacement of the factor $A  = 2\pi  n\sigma^{2}\,[1+ H(1)]$ by $2\pi  n\sigma^{2}= 8\xi $.
The inequality (\ref{condition2}) becomes then 
\be
\label{condition3}
  8\xi > 15.12, \;\;\;\;\; {\rm or}  \;\;\;\;\; \xi > 1.89
\ee
representing for $\xi$ a physically impossible condition 
(beyond close packing). The ring approximation is unable to describe a qualitative change in the hard disk correlation function. For any accessible surface fraction it predicts exponentially damped oscillations.

\subsection{Structural transition in a hard sphere fluid}

The integral equation satisfied by the correlation function $H(x)$ of a three-dimensional hard sphere fluid for $x \gg 1$ has been analyzed 
in Ref.~15 within the ring approximation. The frequencies of the exponential modes $\exp(\kappa x)$  describing the long distance decay of correlations are in this case solutions of the equation (see eq. (22) in \cite{r15})
\be
\label{HSmodes}
\kappa^{2} = 4n\pi\sigma^3 \left( \frac{{\rm sh}(\kappa)}{\kappa}- {\rm ch} (\kappa)\right)
\ee
In passing to the Kirkwood superposition approximation we need only to replace the factor  $4n\pi\sigma^3$ in the above equation by $\lambda = 4n\pi\sigma^3 [1 + H(1)]$. This fact follows from the previously made remark that the ring approximation represents exactly the linearized Kirkwood theory.
In order to determine the range of values of $\lambda > 0$ beyond which the exponential damping becomes impossible we explore the possibility of vanishing of the real part $a$ of the complex number $\kappa=a+i b$. Equation (\ref{HSmodes}) reduces then to
\be
\label{transition1}
b^{2} = \lambda \left(  \cos b - \frac{\sin b}{b} \right), \;\;\; {\rm or}   \;\;\;\;  \lambda = \frac{b^3}{b\cos b - \sin b}
\ee
The absolute minimum of the function
\be
\label{transition2}
y(b) = \frac{b^3}{b\cos b - \sin b}
\ee
in the region where $y(b) > 0$ equals
\be
\label{transition3}
y_{min} = 34.81 ...
\ee
Hence, for $\lambda > 34.81$  equation (\ref{HSmodes}) acquires purely imaginary solutions and the exponential damping vanishes. 
It is quite remarkable that sixty years ago  J.G. Kirkwood, E.K. Maum and B.J. Alder~\cite{r6} concluded from numerical studies of the integral equation for $H(x)$ that when $\lambda$ exeeds $34.8$ the correlation function $H(x)$ is not integrable any more. 
Our method provides a simple analytic confirmation of this result.   

\subsection{The hard rod fluid: testing the method}

 The rigorous calculation of the two-particle correlation function for hard rods (see e.g. Ref.~17) shows that its structure corresponds to exponentially damped oscillations at all possible densities. One finds
\be
\label{RIGH}
n\sigma [H(x) + 1] = \zeta \sum_{k=0}^{\infty}\theta[x-(k+1)] \frac{\zeta^k[x-(k+1)]^k}{k!} \exp\{ -\zeta [ x-(k+1) ] \}
\ee
where 
\[\zeta = \frac{n\sigma}{1-n\sigma} \]

\begin{figure}
\center\includegraphics[width=5in]{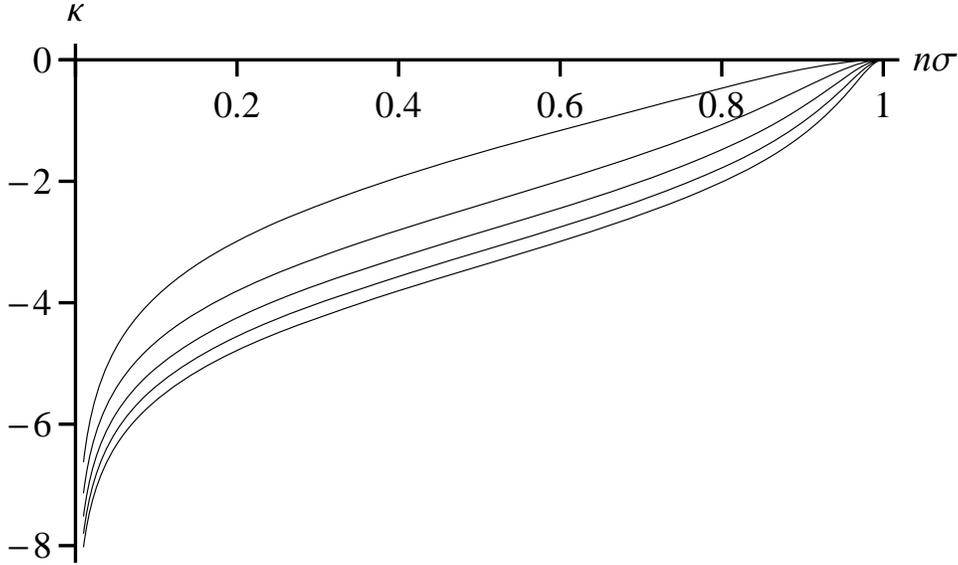}
\caption{The real part of $\kappa_m = -\zeta + W_m( \zeta\exp\zeta )$, m=1,\dots,5 (top to bottom) as a function of $n\sigma = \zeta/(1+\zeta)$.}\label{1d}
\end{figure}

In fact, the superposition law turns out to be exact for a one-dimensional hard rod fluid~\cite{r17}, and the second equation of the
equilibrium hierarchy takes a particularly  simple form
\be
\label{HR1}
   H'(x) = n\sigma [\theta(x-2)H(x-1) - H(x) - \theta(2-x)][H(1)+1]
\ee
The formula (\ref{RIGH}) represents the solution of (\ref{HR1}).
When $x>2$, we find
\be
\label{HR2}
   H'(x) = n\sigma [H(x-1) - H(x)][H(1)+1]
\ee
Applying the same method as that used for hard disks we look for exponential modes
$\exp(\kappa x)$ solving (\ref{HR2}).  The complex frequency $\kappa$ satisfies the equation
\be
\label{modes}
\kappa = n\sigma [H(1)+1]( e^{-\kappa} - 1 ) = \zeta ( e^{-\kappa} - 1 ) 
\ee
It turns out that all solutions of equation (\ref{modes}) can be expressed in terms of the multivalued Lambert W function.
Indeed, the special function $W(z)$ is defined on the complex plane by the equation
\be
\label{LW}
z = W(z)\exp[W(z)] 
\ee
But (\ref{modes}) can be rewritten as
\be
\label{modes2}
(\kappa + \zeta )\exp(\kappa+\zeta)= \zeta\exp\zeta
\ee
 It follows that 
\be
\label{HRmodes}
\kappa = -\zeta + W( \zeta\exp\zeta )
\ee
The principal branch of Lambert function, $W_0(z)$, obeys $W_0(x e^x)=x$ for real $x$, thus $\kappa=0$ for that branch. However, other branches, $W_m(z)$ with $m>0$, give values of $\kappa_m$ with a negative real part, corresponding to the decay of the the correlation function. The first five solutions, $\kappa_m(\zeta), m=1,\dots,5$, are presented in Fig.~\ref{1d} as functions of $n\sigma = \zeta/(1+\zeta)$ (modes with larger $m$ decay faster).

The negative real part of $\kappa_m, \ m>0$  is different from zero for any accessible density, and vanishes only at close packing  where  $n \sigma =1$. 
There are thus no purely imaginary solutions $\kappa=ib$ of equation (\ref{modes}). 
This confirms the correctness of the method, showing the absence of
 any structural transition in one dimension over the whole range of densities below close packing.

\begin{figure}
\center\includegraphics[width=5in]{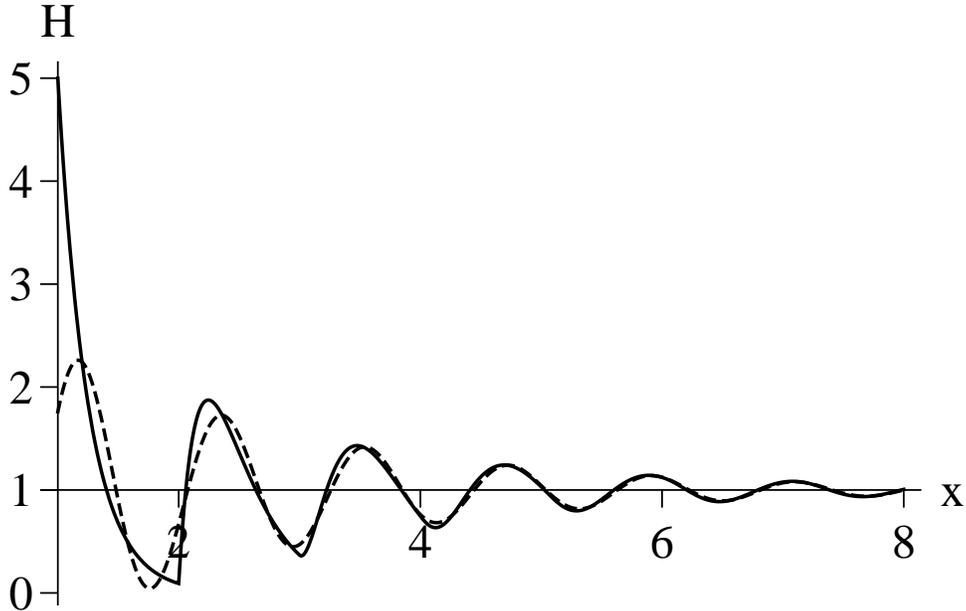}
\caption{The exact form of the correlation function for hard-rod fluid (solid) at $n \sigma=0.8$ and its asymptotic form $H(x)=A e^{\kappa x}$ with $\kappa(\zeta)$ calculated using Eq.~\eqref{HRmodes}.}\label{comp1}
\end{figure}

\begin{figure}
\center\includegraphics[width=5in]{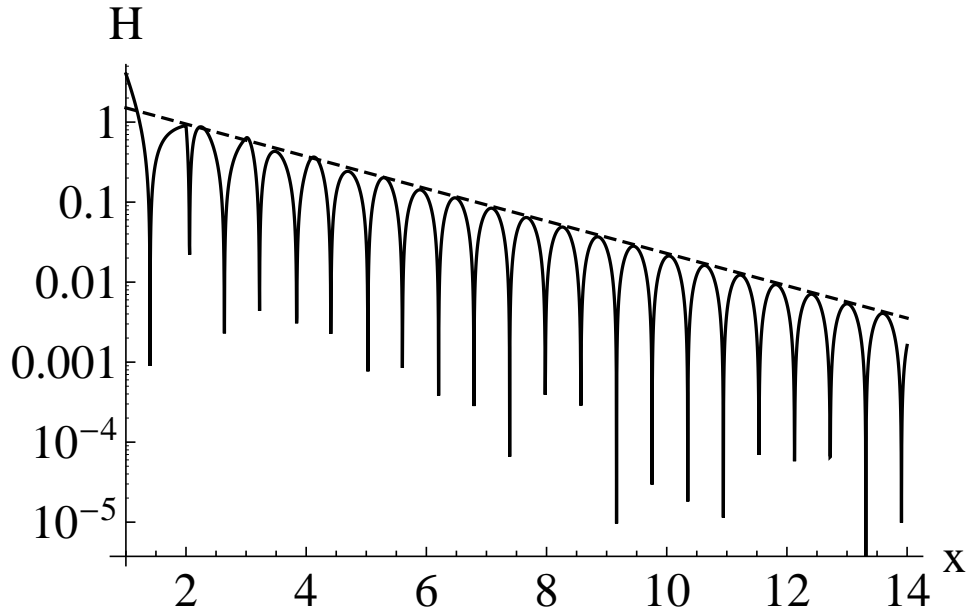}
\caption{The absolute value of $H(x)-1$ for hard rod fluid (solid) at $n \sigma=0.8$ and the exponential asymptote 
$H(x)=A e^{\text{Re}(\kappa) x}$}\label{comp2}
\end{figure}

As illustrated in Figs.~\ref{comp1} and \ref{comp2}, the correlation function $H(x)$ rapidly approaches the asymptotic form given by the slowest decaying mode, $A e^{\kappa_1 x}$. Note that not only the exponential decay rate but also the oscillation period of the function agree with that given by $A e^{\kappa_1 x}$ starting from $x \approx 3 \sigma$.   This shows that the other modes play a negligible role in influencing the behavior of $H(x)$ for intermediate and large $x$ values, thus lending further support to our approach of  focusing on the slowest decay mode only.

\section{Discussion and Conclusions}

The question of whether a system of particles interacting via a purely repulsive potential
(only) can undergo a phase transition has been under continuous investigation since first
posed and addressed by Kirkwood over 70 years ago. For a system of hard disks, the first numerical evidence was provided by Alder and coworkers \cite{r21,r22}. The data reported in Ref.~21 showed that the hard disk freezing transition occurred at a density smaller than the density of closest packing [corresponding to an area fraction of  $\xi_0 = \pi/\sqrt{12} = 0.90690$],  and  suggested that the liquid to solid transition was first order. The most recent Monte Carlo simulations (on a system of $4 \cdot 10^6$ disks) of  Mak \cite{r16} suggest  that melting consists of a continuous transition from the ordered solid to an intermediate (hexatic) phase \cite{r23,r24,r25,r26,r27}  at a packing fraction $\eta = 0.723$, and either a very weak first-order or a continuous transition from the intermediate phase to the fluid phase at a packing fraction $\eta = 0.699$.

From the analysis and numerical evidence presented in Section 4.B, we have calculated
the area fraction $\xi$ at which a structural change in the hard disk fluid can take place, viz.,  $\xi^* \sim 0.622$.  Converting this area fraction to a packing fraction gives  $\eta^*=\xi^*/\xi_0 \sim 0.718$.  When compared to the estimate reported by Mak for the transition from the ordered solid phase to a dense fluid phase,  $\eta \sim 0.723$, one finds the two values are in substantial agreement.  Also of interest is our prediction of a structural transition in a system of hard spheres.  As noted in Section 4.C, Kirkwood, Maum and Alder \cite{r6} found that for values of $\lambda \geq 34.8$, no solutions of the YBG and Kirkwood integral equations exist for which, in their notation,  $x^2[g(x) - 1]$ is integrable.  We find that the value of $\lambda$ beyond which exponential damping becomes impossible is  $\lambda = 34.81$.  Hence, the results for hard spheres appear to be in exact agreement. When we consider a system of hard rods, the analytic method developed here confirms the absence of any structural transition in d=1 for the entire range of densities below close packing  (a result which was already  known  to  Rayleigh \cite{r28} and Korteweg \cite{r29}).

Our prediction of a structural phase transition is based on the analysis of an integral equation whose derivation assumed sufficiently fast decay of correlations.  If  a phase becomes ordered and correlations do not decay,  the integral equation to which our method was applied does not hold.  This can preclude the possibility of studying a region where a new equilibrium phase may be formed.  In particular, the identification of a region intermediate between melting and freezing (see following paragraph)  and the characterization of an ordered (solid) phase is certainly beyond the scope of our approach.  To study these questions within the Kirkwood superposition approximation, one must go back to the original YBG hierarchy equation (Kirkwood's closure does not assume the rapid decay of correlations).

To develop this point further, there are three structural aspects of the hard disk transition
that are not captured by the method developed in this paper. 
 First, we find no evidence for the existence of  an
intermediate, or hexatic, phase predicted by the Kosterlitz, Thouless, Halperin, Nelson
and Young  (KTHNY) theory of d=2 melting \cite{r23,r24,r25,r26,r27},  and supported by  Mak's  Monte Carlo simulations. Second, we find no evidence for the development of a shoulder on the second maximum of disk radial distribution function  in the vicinity of  the freezing 
transition ($\eta= 0.686$),  reported  by Truskett et al. \cite{r30} based on molecular dynamics simulations, and later correlated with structural rearrangements occurring at increasing disk density  \cite{r31,r32,r33,r34}. Thirdly, we cannot confirm the existence of regions of five-fold coordination in the dense fluid phase, first predicted by Bernal \cite{r35,r36,r37} based on his "ball and spoke" model of a random assembly of  hard-core particles, although it has been conjectured that the hexatic phase might be correlated with randomly dispersed regions of 5-, 6-, and 7-member disk clusters forming percolated tessellations that span the d=2 space \cite{r38}.
 
The larger point, however, relates to the original Kirkwood prediction, viz. that a system of particles interacting via purely {\it repulsive} forces, here hard disks but also hard spheres, can undergo a phase transition.  Although not widely accepted at first, following the work of Onsager on the isotropic-nematic transition in a  d=3 dimensional system of thin hard rods \cite{r39}, there developed a gradual realization that a phase transition can be entropy driven.  As elaborated by Frenkel \cite{r40}, in hard-core systems the entropy in the ordered phase is actually larger than the fluid phase; quoting directly,  ``the entropy {\it decreases} because the density is no longer uniform in orientation or position, but the entropy {\it increases} because the free-volume per particle is larger in the ordered than in the disordered  phase.''   The present contribution provides further evidence for the essential correctness of Kirkwood's insight.

\appendix

\section{Derivation of Eq.(5) from the BBGKY hierarchy} 

Consider a gas of hard disks of mass $m$ and diameter $\sigma$. We denote by $j \equiv (\br_{j},\bv_{j}) $, $j=1,2, ... $ the  one-particle state  in which a disk has position $\br_{j}$ and velocity $\bv_j$. The average number density of $s$-particle clusters occupying at time $t$ the s-particle state $(1,2, ... ,s)$ is called the $s$-particle reduced distribution 
$f_{s}(1,2, ... ,s;t)$.

The dynamical evolution of the hard disk fluid is described in the thermodynamic limit by the BBGKY hierarchy equations.  The  second of them establishes a relation between $f_{2}$ and $f_{3}$
\be
\label{BBGKY}
\left(\derpar{}{t}+\VEC{v}_1\cdot\derpar{}{\VEC{r}_1} +
\VEC{v}_2\cdot\derpar{}{\VEC{r}_2} -\T(1,2) \right)\ff(1,2;t)=\\
\int d3\, (\T(1,3)+\T(2,3))\fff(1,2,3;t)
\ee
The effects of binary collisions are described by the operator $\T(i,j)$  
\be
\label{collop}
\T(i,j)=\sigma \int d\s\,\vs{1}{2}\,\theta(\vs{i}{j})\left[\delta(\VEC{r}_{ij}-\si )b_{\s}
-\delta(\VEC{r}_{ij}+\si )\right],
\ee
Here the Dirac $\delta $-distributions restrict the distances $|\VEC{r}_{ij}|$ between the centers of the disks at the moment of impact to their diameter $\sigma = |\si |$.
The vector $\si = \sigma\s $ is oriented perpendicularly to the surface of colliding disks at the point of impact. 
The action of the operator $b_{\s}$ consists in replacing the velocities $\bv_{i},\bv_{j}$ by their precollisional values $\bv_{i}',\bv_{j}'$ corresponding to the inverse elastic collision. As in elastic collisions the kinetic energy is conserved, in the case of products of Maxwell distributions 
\[ \phi(v) = \left( \frac{m}{2\pi k_{B}T } \right)\exp(-mv^{2}/2k_{B}T) \] 
we find 
\be
b_{\s}[\phi(\bv_{i})\phi(\bv_{j})] = \phi(\bv_{i}')\phi(\bv_{j}')=\phi(\bv_{i})\phi(\bv_{j})
\ee
Hence,  in the case of equilibrium reduced distributions  
\be
\label{equilibrium}
f_{s}(1,2, ... ,s) = n_{s}(\br_{1}, \br_{2} ... \br_{s})\phi(\bv_{1})\phi(\bv_{2}) ... \phi(\bv_{s})
\ee
equation (\ref{BBGKY}) takes the form
\be
\label{YBGi}
\left( \bv_1\cdot\derpar{}{\br_1} +
\bv_2\cdot\derpar{}{\br_2} -  \sigma \int d\s\,\vs{1}{2}\,\delta(\br_{ij}-\si ) \right) n_{2}(r_{12})\phi(\bv_{1})\phi(\bv_{2}) =
\ee
\[ \sigma \int d\br_{3}\, d\bv_{3}\,
\int d\s\, \left[ \vs{1}{3}\, \delta(\br_{13}-\si ) +\vs{2}{3}\, \delta(\br_{23}-\si )\right]\]
\[\times  n_{3}(r_{12}, r_{13}, r_{23})\phi(\bv_{1})\phi(\bv_{2})\phi(\bv_{3}) \]
We can divide both sides of equation (\ref{YBGi}) by the product $\phi(\bv_{1})\phi(\bv_{2})$, and perform integration over the $\bv_{3}$ variable.  Moreover, we introduce explicitly the excluded volume factors by using equations (\ref{n2}), (\ref{n3}).
The hierarchy equation becomes
\be
\label{YBGii}
\left( \bv_{12}\cdot\derpar{}{\br_{12}} 
 -  \sigma \int d\s\,\vs{1}{2}\,\delta(\br_{12}-\si ) \right) \theta(r_{12}-\sigma)y_{2}(r_{12})  =
\ee
\[ n \sigma \theta(r_{12}-\sigma) \int d\br_{3}\,
\int d\s\, \left[ \bv_{1}\cdot\s \, \delta(\br_{13}-\si )\theta(r_{23}-\sigma)-\si )y_{3}(r_{12},\sigma ,r_{23}) \right. \]
\[\left.  + \bv_{2}\cdot\s \, \delta(\br_{23}-\si ) \theta(r_{13}-\sigma) 
y_{3}(r_{12},r_{13},\sigma) \right] \]
We now use the identity
\be
\label{identity}
\bv_{12}\cdot\derpar{}{\br_{12}}\theta(r_{12}-\sigma)\equiv \sigma \int d\s\,\vs{1}{2}\,\delta(\br_{12}-\si ) 
\ee
which reduces the left hand side of (\ref{YBGii}) to
\be
\label{Left}
L = \theta(r_{12}-\sigma)\bv_{12}\cdot\derpar{}{\br_{12}}y_{2}(r_{12}) 
\ee
On the right-hand side owing to the presence of $\delta $-distributions we can perform integration over variable $\br_{3}$ thus obtaining
\be
\label{Right}
R = n \sigma \theta(r_{12}-\sigma)\int d\s\,  \bv_{12}\cdot\s \,\theta(|\br_{12}-\si )|-\sigma )
y_{3}(r_{12},\sigma , |\br_{12}-\si )|) 
\ee
As the equality $L=R$ must hold for any value of the relative velocity $\bv_{12}$, we finally find 
 (when $r_{12}\ge\sigma$)
\be
\label{YBGfinal}
\frac{d y_{2}(r_{12})}{dr_{12}} = n \sigma \int d\s\,  \hat{\br}_{12}\cdot\s \,\theta(|\br_{12}-\si |-\sigma )
y_{3}(r_{12},\sigma , |\br_{12}-\si |) 
\ee
  which is equation (\ref{YBGy}) of section 2.

\section{Derivation of Eq.(27)}
We perform here the angular integration in the contribution to the correlation function
\be
\label{A1}
I^{**}(x)= - \frac{1}{2} \int_{x}^{\infty} dz\,\int_{0}^{2\pi }d\phi \, \cos\phi\, \theta( z-2\cos\phi )H(\sqrt{z^{2}+1-2z\cos\phi})
\ee
\[ = -  \int_{x}^{\infty}dz \int_{0}^{\pi/2 }d\phi \, \cos\phi 
[  \theta(z-2\cos\phi)H(\sqrt{z^{2}+1-2z\cos\phi}) - H(\sqrt{z^{2}+1+2z\cos\phi})]  \]

Changing the order of integrations with the use of the asymptotic decay of the correlation function we arrive at a convenient formula
\be
\label{A2}
I^{**}(x)= - \int_{0}^{\pi/2 }d\phi \, \cos\phi \int_{x-\cos\phi}^{x+\cos\phi}ds\, \theta(s-\cos\phi)
H(\sqrt{s^{2}+\sin^{2}\phi})
\ee
Putting then $\mu = \sin\phi$ we get
\be
\label{A3}
I^{**}(x) =- \int_{0}^{1}d\mu \, \int ds \theta(x+\sqrt{1-\mu^2}-s)\theta(s-x+\sqrt{1-\mu^2})
\ee
\[\times  H(\sqrt{s^2 + \mu^2})\theta(s-\sqrt{1-\mu^2})   \]

We now introduce a new integration variable
\[ z=\sqrt{s^2 + \mu^2}  \]
As $zdz = sds$ we find
\be
\label{A4}
I^{**}(x) =  -\int_{0}^{1}d\mu \, \int dz \ \frac{z}{\sqrt{z^2-\mu^2}}H(z)\theta(\sqrt{z^2-\mu^2}-\sqrt{1-\mu^2})\theta(\sqrt{1-\mu^2}-
|x-\sqrt{z^2-\mu^2}|)
\ee
\[ =-  \int dz \ zH(z)\theta(z-1) \int_{0}^{1}d\mu \frac{1}{\sqrt{z^2-\mu^2}}\theta(\sqrt{1-\mu^2}-|x-\sqrt{z^2-\mu^2}|) \]

Here the inequality $\sqrt{1-\mu^2}>|x-\sqrt{z^2-\mu^2}|$ is equivalent to 
\[ \sqrt{z^2-\mu^2}> \frac{x^2+z^2-1}{2x}  \] 
which leads to the formula
\be
\label{A5}
I^{**}(x) = -\int dz \ zH(z)\theta(z-1) \int_{0}^{1}d\mu \frac{1}{\sqrt{z^2-\mu^2}}\theta(\sqrt{z^2-\mu^2}- \frac{x^2+z^2-1}{2x} )
\ee
Putting $\mu = z\nu $ we find
\be
\label{A6}
I^{**}(x) =- \int dz \ zH(z)\theta(z-1) \int_{0}^{1/z}d\nu  \frac{1}{\sqrt{1-\nu^2}}\theta(\sqrt{1-\nu^2}- \frac{x^2+z^2-1}{2xz} )
\ee
The change of the integration variable $w=\sqrt{1-\nu^2}$ yields the formula
\be
I^{**}(x) =  -\int dz \ zH(z)\theta(z-1)\int_{\sqrt{1-1/z^2}}^{1}\frac{dw}{\sqrt{1-w^2}}\theta( w - \frac{x^2+z^2-1}{2xz} )                                                                     
\ee
\[ = -  \int dz \ zH(z)\theta(z-1)\int_{(x^2+z^2-1)/2xz}^{1}\frac{dw}{\sqrt{1-w^2}}\theta( 1 - \frac{x^2+z^2-1}{2xz} )    \]
\[  = -\int dz \ z H(z) \theta( z-1) \theta[1- (x-z)^2] \left\{   \frac{\pi}{2} - \arcsin\left( \frac{x^{2}+z^{2}-1}{2xz} \right)  \right\} \]
Finally, we arrive at the result 
\be
I^{**}(x) = - \int_{x-1}^{x+1} dz \ z Y(z) \theta( z-1) \arccos\left( \frac{x^{2}+z^{2}-1}{2xz}\right)  
\ee
used in equation (\ref{**}).

\end{document}